\documentclass[aps,prl,reprint,groupedaddress]{revtex4-1}
\usepackage{graphicx}

\begin{document}

\title{Adaptive walks and extreme value theory}

\author{Johannes Neidhart$^1$ \& Joachim Krug$^1$}

\affiliation{$^1$Institute for Theoretical Physics, University of Cologne, 50937 K\"oln, Germany}

\date{\today}

\begin{abstract}
We study biological evolution in a
high-dimensional genotype space in the regime of rare mutations and
strong selection. The population performs an uphill walk
which terminates at local fitness maxima. Assigning
fitness randomly to genotypes, we show that the mean walk length
is logarithmic in the number of initially available
beneficial mutations, with a prefactor determined by  
the tail of the fitness distribution. This result is derived
analytically in a simplified setting where the mutational neighborhood
is fixed during the adaptive process, and confirmed by numerical
simulations.

\end{abstract}






\maketitle

The adaptation of a population to a novel environment is a fundamental process of evolutionary
biology which continues to attract considerable attention from theoretical \cite{Orr2005} as well as 
experimental \cite{Elena2003}
perspectives. Adaptation is driven by the occurrence of mutations that
are beneficial
in the new environment and therefore spread in the population, leading to an increase of 
fitness over time. This process displays a variety of dynamical patterns \cite{Park2010} that depend on the
supply of beneficial mutations (governed by the product of population size $M$ and
mutation rate $U$) as well as on the structure of the \textit{fitness landscape}, which encodes
how the genetic configuration of an organism (its \textit{genotype}) affects the number of offspring
it will leave in the next generation. 

A particularly simple, yet biologically relevant limit of adpative dynamics is the regime
of strong selection and weak mutation (SSWM), where mutations are sufficiently rare to 
be treated as independent events, $MU \ll 1$, and selection is strong enough for deleterious
mutations (which decrease fitness) to be unable to spread \cite{Gillespie1983,Gillespie1984,Orr2002}. 
In the SSWM regime the population
is genetically homogeneous most of the time, and its dynamics can be described by a 
point in the space of genotypes which performs an \textit{adaptive walk} towards higher
fitness. Because of the low mutation rate such a walk is constrained to move by single
mutational steps, and it terminates when a local fitness maximum is reached, where no
nearest neighbor genotypes are available that would confer higher fitness.
Despite its strongly simplified nature, the
adaptive walk model is in principle amenable to quantitative tests
in microbial evolution experiments 
\cite{Rokyta2005,Schoustra2009,Rokyta2009,Miller2011}.

In the present Letter we study the length of such adaptive walks in a simple  
model of a rugged fitness landscape, where fitness values $F_i$ of genotypes $i$ 
are assumed to be independent
random variables drawn from a common probability density $\rho(F)$. 
The genotype space is a generalized hypercube formed by 
sequences of $L$ letters drawn from an alphabet of size $a$, such that each
genotype has $N=(a-1)L$ single mutant neighbors \cite{Kauffman1987}. 
The walk is then specified by the transition probability $P_{ij}$ from genotype $i$ to a neighboring
genotype $j$ of higher fitness, $F_j > F_i$. In the SSWM regime $P_{ij}$
is proportional to the \textit{fixation probability} of the corresponding beneficial mutation, i.e.
the probability that it will become dominant rather than going extinct
due to demographic fluctuations \cite{Kimura1962,Patwa2008}. When the fitness
difference $\Delta F_{ij} = F_j - F_i$ between the initial and final genotype  is 
small in absolute terms, $\vert \Delta F_{ij} \vert \ll 1$, while
still maintaining the strong selection condition $M \vert \Delta F_{ij} \vert \gg 1$, , 
the fixation probability is proportional to $\Delta F_{ij}$, and normalization 
leads to the expression \cite{Gillespie1983,Gillespie1984,Orr2002}
\begin{equation}
\label{Pij}
P_{ij} = \frac{\Delta F_{ij}}{\sum_{k:F_k > F_i} \Delta F_{ik}}.
\end{equation} 
After the transition the population has fitness $F_j$ and encounters a new set of 
random fitness values (apart from the fitness $F_i$ of the preceding genotype,
which is however inaccessible because $F_i < F_j$). 

Assuming that $n$ fitter neighboring genotypes are available at the starting point of the adaptive walk,
we ask for the mean number of steps $\ell(n,N)$ that are required to reach a local fitness
maximum. Since most mutations available to a viable genotype are expected to be deleterious or neutral \cite{Eyre2007},
we are mainly interested in the behavior of $\ell$ when $N \gg n \gg 1$. 
Simplified variants of this problem have been considered in previous work. In the \textit{random}
adaptive walk the dependence of the transition probability on fitness differences is ignored, and
all available fitter neighbors are chosen with equal probability, which leads to 
$\ell_\mathrm{random} \approx \ln n + c_\mathrm{random}$ with $c_\mathrm{random} \approx 1.1$
\cite{Kauffman1987,Macken1989,Flyvbjerg1992}.
On the other hand, for \textit{greedy} walks which always move to the neighboring
genotype of highest fitness, the walk length remains finite for $N, n \to \infty$ and 
attains a limiting value of $\ell_\mathrm{greedy} = e -1 \approx 1.71$ \cite{Orr2003}. 
 
For the full problem defined by the fitness-dependent transition probability
(\ref{Pij}) we show below that the asymptotic behavior of the mean walk length is generally logarithmic, 
with a coefficient that depends on the form of the 
tail of the fitness distribution $\rho(F)$. According to extreme value theory (EVT), the tail can be 
represented by the generalized Pareto form \cite{Pickands1975,Beisel2007,Joyce2008,Note1}
\begin{equation}
\label{Pareto}
\rho(F) = \left(1 + \kappa F \right)^{-\frac{\kappa + 1}{\kappa}}
\end{equation}
where the shape parameter $\kappa$ serves to distinguish between the
different universality classes of EVT \cite{deHaan}. For $\kappa > 0$ the density (\ref{Pareto}) is defined
for all $F > 0$ and decays as a power law, representing the Fr\'echet class of EVT, whereas for $\kappa < 0$ its support
is restricted to the interval $[0,-\kappa^{-1}]$ and the distribution belongs to the Weibull class. The Gumbel class,
comprising distributions of unbounded support that decay faster than a power law, is recovered in the limit $\kappa \to 0$.
In previous work \cite{Joyce2008} it has been shown that the adaptive walk with fitness distribution (\ref{Pareto}) reduces 
to the random (greedy) limit for $\kappa \to - \infty$ ($\kappa \to \infty$). 
For $\kappa \to - \infty$ the density (\ref{Pareto}) develops a 
$\delta$-function singularity at the upper boundary of its support, which implies that all available mutants have the same fitness
and (\ref{Pij}) reduces to a random choice. On the other hand, for $\kappa \to \infty$ the density (\ref{Pareto}) 
becomes extremely broad, such that the fitness of the most fit mutant in a neighborhood is typically much larger than all other fitness
values and (\ref{Pij}) reduces to the greedy rule.

In terms of the parametrization (\ref{Pareto}), our main result for the mean walk length reads
\begin{equation}
\label{beta}
\ell \approx \beta \ln n \;\; \textrm{with} \;\; \beta =
\frac{1-\kappa}{2-\kappa} \;\; \textrm{for} \;\; \kappa \leq 1.
\end{equation}
This expression recovers the random limit ($\beta = 1$) for $\kappa \to - \infty$, and shows that 
the greedy limit ($\beta = 0$) is attained at $\kappa = 1$, where the density (\ref{Pareto}) ceases to have a finite
first moment.  The result $\beta = 1/2$ for the Gumbel class 
was previously obtained numerically by Orr \cite{Orr2002}
(see below), and analytically by Jain and Seetharaman \cite{Jain2011} using an approach along the lines of \cite{Flyvbjerg1992}. 
Surprisingly, the expression (\ref{beta}) also appears in the context
of a completely different evolution model of quasispecies type, which applies 
in the limit of infinite populations \cite{Krug2003,Jain2005,Sire2006}. 
The reason for this coincidence will be discussed at the end of the paper.    

\begin{figure}[ht]\begin{center}\includegraphics[width=0.4\textwidth]{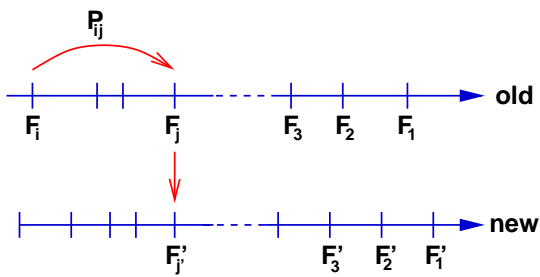}\end{center}
\caption{(Color online) Illustration of the two processes involved in
  a step of the adaptive walk. Starting from a genotype of fitness
  rank $i$ in its current mutational neighborhood (upper fitness axis), the population
  moves to rank $j < i$ with probability $P_{ij}$. In the new
  neighborhood (lower fitness axis) the rank of the current genotype
  is $j'$. In the Gillespie aproximation the old and the new
  neighborhoods are the same. \label{Fig:oldnew}} \end{figure}

\noindent
\textit{The Gillespie approximation.} Our analysis is based on an approximation first introduced by 
Gillespie \cite{Gillespie1983}. The key
idea is to ignore the change 
in available fitness values that occurs after a jump
of the adaptive walk, which implies that the entire adaptive process
proceeds in a single, fixed neighborhood
(Fig. \ref{Fig:oldnew}). 
The expected length of the walk is then equal to the first passage time (or absorption time) 
of the Markov chain defined by the transition probability (\ref{Pij}) for a fixed set of fitness values
$F_k$. For the following discussion it will be convenient to label the fitness values by their
rank, such that $F_1 > F_2 > ... > F_N$. The mean absorption time to the final state of maximal fitness
$F_1$, starting from fitness rank $n$, is then given by \cite{Gillespie1983}
\begin{equation}
\label{absorption}
t_n = H_{n-1} - \sum_{i=1}^{n-1} \frac{\lambda_i}{\lambda_{n} (n-1)} - \sum_{i=1}^{n-1} \sum_{j=i+1}^{n-1}
\frac{\lambda_i}{\lambda_j j (j-1)} 
\end{equation}
where $H_k = \sum_{i=1}^k \frac{1}{i}$ is the $k$th harmonic number, and  
\begin{equation}
\label{lambda}
\lambda_i = \sum_{k=1}^{i-1} k (F_{k}-F_{k+1}) = \sum_{k=1}^{i-1} k \Delta_k
\end{equation}
with $\lambda_1 = 0$ and \textit{fitness gaps} $\Delta_k = F_{k}-F_{k+1}$. Because fitness only increases during the process,
the absorption time is obviously independent of the fitness values $F_{n+1}, F_{n+2},...,F_N$ above the 
starting rank.

Within the Gillespie approximation, the adaptive walk length $\ell$ is obtained by averaging the absorption time 
(\ref{absorption}) with respect to the fitness distribution $\rho(F)$. Gillespie observed that
the problem simplifies significantly if $\rho(F)$ is assumed
to fall into the Gumbel universality class of EVT. Taking the limit $N \to \infty$ at fixed $n$, the 
$n$ superior fitness values lie in the tail of the distribution, and it is known that the scaled fitness
ranks $k \Delta_k$ converge to independent, identically distributed exponential random variables \cite{deHaan}. 
It then  follows by symmetry that the average ratios in (\ref{absorption}) are $\langle \frac{\lambda_i}{\lambda_j} \rangle = \frac{i-1}{j-1}$,
and evaluation of the sum yields the simple result \cite{Gillespie1983,Orr2002}
$\langle t_n \rangle = \frac{1}{2} (H_{n-1} + 1) \approx \frac{1}{2} \ln n + \frac{1}{2} (\gamma + 1)$,
where $\gamma \approx 0.577215...$ denotes Euler's constant.
Simulations of the full problem show that the mean walk length differs from this approximate result
only by an offset in the constant correction term, which is given by $c_{0} \approx \frac{1}{2} (\gamma + 1) + 0.44$
\cite{Orr2002}. A similar calculation for the model with random choice of fitter neighbors yields a mean absorption time
of $\langle t_n \rangle = H_{n-1} \approx \ln n + \gamma$ \cite{Orr2002}, which again differs from the mean walk length of the full 
model \cite{Macken1989,Flyvbjerg1992} (quoted above) only by a small shift in the constant term. We will show below
that the close agreement between the Gillespie approximation and the full model extends to general
fitness distributions, and provide a qualitative explanation for this behavior. 

\noindent
\textit{General fitness distributions.} We now turn to the approximate evaluation of the absorption time (\ref{absorption}) for the other 
EVT classes. As a representative of the Fr\'echet class we choose the Pareto distribution $\rho(F) = \mu F^{-(\mu + 1)}$, $F \geq 1$, 
which is a shifted and rescaled version of (\ref{Pareto}) with $\mu = 1/\kappa$. A straightforward calculation shows that the
expected value of the $k$th out of $N$ fitness values is given by 
\begin{equation}
\label{Pareto_mean}
\langle F_k \rangle = \frac{\Gamma(N+1) \Gamma(k-\frac{1}{\mu})}{\Gamma(N+1-\frac{1}{\mu}) \Gamma(k)} \approx \left( \frac{N}{k} \right)^{\frac{1}{\mu}}
\end{equation}
for $N \gg k \gg 1$. To estimate the fitness gap we take the derivative with respect to $k$ \cite{Note2}, $\langle \Delta_k \rangle \approx -\frac{\partial}{\partial k} 
\langle F_k \rangle \sim N^{\frac{1}{\mu}} k^{-1-\frac{1}{\mu}}$. Approximating the sum in (\ref{lambda}) by an integral we then find 
$\lambda_i \sim N^{\frac{1}{\mu}} i^{1-\frac{1}{\mu}}$, and hence
$\lambda_i/\lambda_j  \sim (i/j)^{1-\frac{1}{\mu}}$.
Inserting this into (\ref{absorption}) and replacing sums by integrals we see that the first sum converges to a constant
for $n \to \infty$, while the second, double sum diverges logarithmically as $\frac{\mu}{2 \mu - 1} \ln n$. Thus to leading order we find
$\langle t_n \rangle \approx \left( 1 - \frac{\mu}{2 \mu - 1} \right) \ln n = \frac{\mu - 1}{2 \mu - 1} \ln n$,
which is identical to (\ref{beta}) with $\kappa = 1/\mu$. 

The calculation for the Weibull class of distributions with bounded support is similar. We consider
distributions on the unit interval of the form $\rho(F) = (\nu+1) (1 - F)^{\nu}$ with $\nu \geq -1$, corresponding to 
(\ref{Pareto}) with $\kappa = -\frac{1}{\nu + 1}$. The mean of the $k$th out of $N$ values drawn from this distribution is 
given by 
$\langle F_k \rangle  
\approx 1 - \left(\frac{k}{N} \right)^{\frac{1}{\nu+1}}$
for $N \gg k \gg 1$, and along the same lines of reasoning used previously we find that $\lambda_i/\lambda_j \sim (i/j)^{\frac{\nu+2}{\nu+1}}$. 
Again, this implies that the first sum on the right hand side of (\ref{absorption}) converges, whereas the second double sum diverges logarithmically,
leading finally to 
$\langle t_n \rangle \approx \left( 1 - \frac{\nu+1}{2\nu+3} \right) \ln n = \frac{\nu+2}{2 \nu +3} \ln n$,
in agreement with (\ref{beta}). The result $\ell \approx \frac{2}{3} \ln n$ for the uniform distribution
($\nu = 0$) was also obtained in \cite{Jain2011}. 

\begin{figure}[ht]
\begin{center}\includegraphics[width=0.45\textwidth]{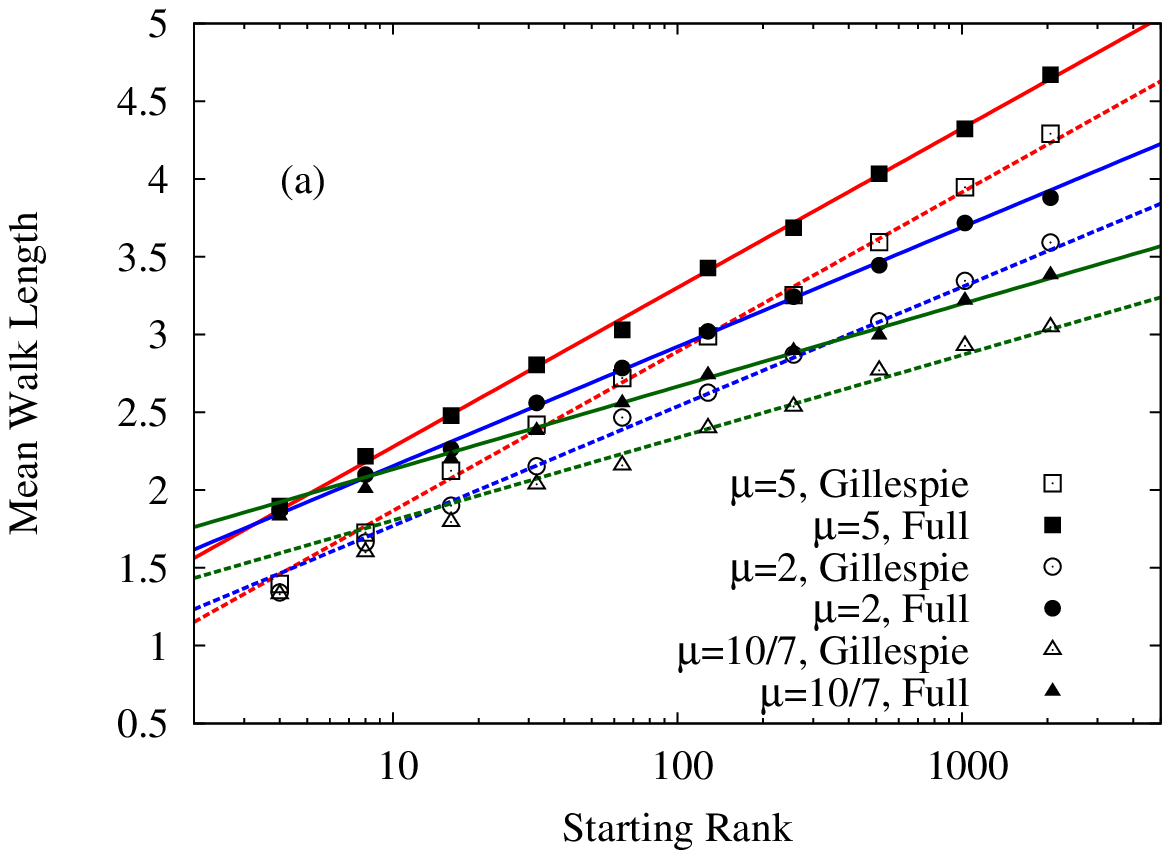}\end{center}
\begin{center}\includegraphics[width=0.45\textwidth]{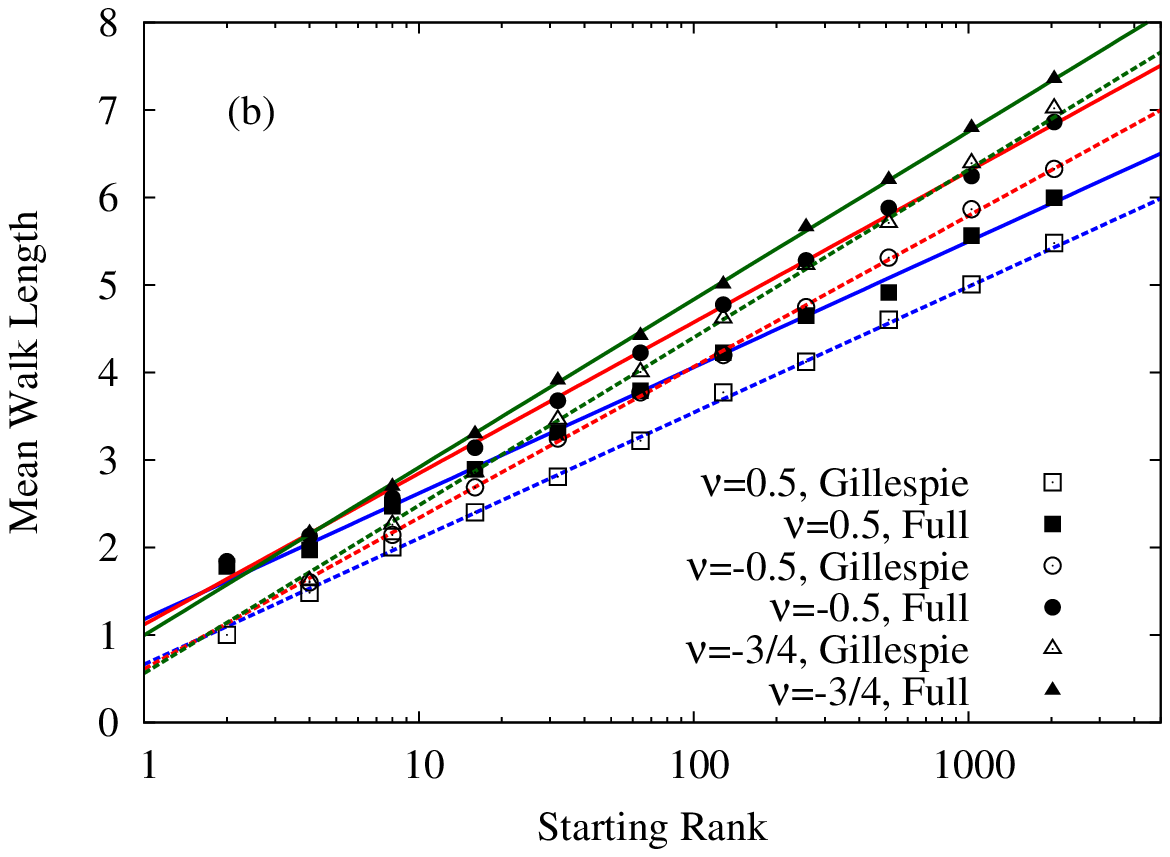}\end{center}
\caption{(Color online) Simulation results for the full adaptive walk
  model (full symbols and lines) and the Gillespie approximation (open
  symbols and dashed lines). Slopes of lines are given by 
  (\ref{beta}) and intercepts have been fitted to the numerical
  data. (a) Fr\'echet class with $\mu = \frac{1}{\kappa} = \frac{10}{7}$, 2 and 5. 
The fitted intercepts are $c_\kappa = c_{7/10} = 1.60$, $c_{1/2} = 1.39$ and $c_{1/5} = 1.25$ for the full model and
$\tilde c_{7/10} = 1.27$, $\tilde c_{1/2} = 1.00$, $\tilde c_{1/5} = 0.84$ for the Gillespie approximation. 
(b) Weibull class with $\nu = -(1 + \frac{1}{\kappa}) = -0.75$, $-0.5$ and $0.5$. 
Fitted intercepts are $c_{-2/3} = 1.18$, $\tilde c_{-2/3} = 0.66$, $c_{-2} = 1.12$, $\tilde c_{-2} = 0.61$, $c_{-4} = 1.00$ and $\tilde c_{-4} = 0.56$ . In all 
cases $c_\kappa > \tilde c_\kappa$. 
\label{Fig:sim1}} \end{figure}

\noindent
\textit{Simulations.} Next we compare the prediction (\ref{beta}) to simulations, using both the full adaptive walk model and
the simplified Gillespie model in a fixed mutational neighborhood. In
the simulations of the full model, we avoided an explicit representation of the genotype space by 
creating the fitness values encountered
during the walk 'on the fly'. This ignores the possibility of the same genotype
being encountered more than once during the walk, which is however negligible for large $N$
\cite{Flyvbjerg1992}. The total size of the neighborhood was $N=4000$ in all
cases, the starting rank was varied from $n=2^2 = 4$ to $n=2^{11} = 2048$ in factors of 2, and results were averaged
over 1000 independent realizations. As can be
seen in Fig.~\ref{Fig:sim1}, the asymptotic prediction (\ref{beta}) is
well satisfied in both kinds of simulations.

To rationalize the observed close agreement between the Gillespie approximation and the full adaptive walk, we analyze the effect that the
two processes involved in a single step of the walk have on the rank of the current genotype (Fig. \ref{Fig:oldnew}). In the first process,
the choice of a fitter neighbor according to the transition probability $P_{ij}$, the rank of the genotype changes by an amount that is 
proportional to the initial rank; to be specific, the expected new rank $j$ conditioned on the original rank $i$ is given by 
$\langle j \rangle =  \frac{1}{2} \beta i$ for $i \gg 1$ \cite{Joyce2008}. The change of rank due to the subsequent
change of the mutational neighborhood (which is omitted in the Gillespie approximation) can be deduced from the classic analysis
of the number of exceedances \cite{Rokyta2006,Gumbel1950}, which shows that the expected new rank $j'$ conditioned on the old rank $j$ is 
$j+1$, with a variance of order $j$. Thus for $i,j \gg 1$ the change in rank due to the change in neighborhood is a small perturbation
(of relative size $\frac{1}{\sqrt{j}}$) of the change that occurs in the first process, which explains the quantitative
accuracy of the Gillespie approximation. The fact that the change of neighborhood on average increases the rank is consistent 
with the numerical observation that the adaptive walks in the full
model are always slightly longer than in the Gillespie approximation
(Fig.~\ref{Fig:sim1}).

\noindent
\textit{Relation to quasispecies models.} The quasispecies approach to
evolution assumes very large populations, $MU \to \infty$, such that
demographic fluctuations are absent and the adaptive process is
completely deterministic \cite{Jain2007a}. In an 
uncorrelated random fitness landscape the most populated genotype then
performs a kind of `adaptive flight', 
which is essentially constrained to move between
local fitness maxima and terminates only when the global fitness maximum is reached \cite{Krug2003,Jain2005}. In the simple case of a \textit{one-dimensional} genotype space, the length of 
such an adaptive flight depends logarithmically on the number of genotypes with a prefactor given precisely
by the expression in (\ref{beta}), a behavior that was first observed numerically \cite{Krug2003} and subsequently derived
analytically in \cite{Sire2006}. The formal relation to the adaptive walk problem can be traced back to the fact that the transition
probability of the adaptive flight, which describes the rate at which the most populated genotype jumps from one fitness peak to the
next, depends linearly on the fitness difference between the two peaks in the same way as the fixation probability (\ref{Pij}) 
\cite{Sire2006}. This structure also appears in the analysis of the collision statistics of a one--dimensional gas with quenched random velocities \cite{Bena2007}. 

Employing a completely different mathematical approach, Sire \textit{et
  al.} \cite{Sire2006} computed the mean length of the adaptive flights
as well as the corresponding variance (see also \cite{Bena2007}). Using their result one finds that 
the index of dispersion $I$ (defined as the ratio of the variance to the mean) depends on the EVT parameter $\kappa$ according to the 
simple expression 
$I = \frac{1+(1-\kappa)^2}{(2-\kappa)^2}$,
which takes its minimal value $I=\frac{1}{2}$ for the Gumbel class
($\kappa = 0$) and approaches unity for $\kappa \to - \infty$ as well
as  for $\kappa \to 1$. 
This formula reproduces the results obtained in \cite{Jain2011} for
$\kappa = 0$ and $\kappa = -1$, and
we have checked numerically that it applies to the full adaptive 
walks problem for general $\kappa$. Thus, while the walk length has a
Poisson distribution in the case of random dynamics \cite{Flyvbjerg1992}, 
in general the fluctuations are sub-Poissonian.

\noindent
\textit{Conclusions.} We have analyzed a simple, paradigmatic model for the evolution of populations subject to rare mutations
and strong selection, and derived a precise asymptotic relation between the length of adaptive walks and the tail of the underlying
fitness distribution. While the predicted asymptotics may be difficult to observe in experiments, the EVT shape 
parameter $\kappa$ can be estimated experimentally \cite{Beisel2007}, 
and examples with $\kappa = 0$ \cite{Kassen2006}, $\kappa < 0$ \cite{Rokyta2008} and $\kappa > 0$ \cite{Schenk} have been identified.

An important restriction of our model is the assumption that fitness values of different genotypes are uncorrelated.   
Indeed, a recent study comparing the distributions of beneficial fitness effects encountered during the first and second steps of an adaptive walk
found strong evidence for fitness correlations between neighboring genotypes \cite{Miller2011}. Such correlations are likely to significantly
affect the results presented here, and will be addressed in the future.

\begin{acknowledgments}

This work was supported by DFG within SFB 680 and the Bonn Cologne
Graduate School of Physics and Astronomy. We thank Kavita Jain and
Henrik Flyvbjerg for useful correspondence.

\end{acknowledgments}

\end{document}